\documentclass[pre,twocolumn,aps,floats,longbibliography,
floatfix,showpacs,showkeys,amsmath,amssymb,nofootinbib]{revtex4-1}

\usepackage{dcolumn}
\usepackage{bm}
\usepackage{txfonts}
\usepackage{pxfonts}
\usepackage{graphicx}
\usepackage{psfrag}
\usepackage{times}
\usepackage{multirow} 

\begin{document}

\title{Logistic modelling of economic dynamics}



\author{Arnab K. Ray}\email{arnab\_kumar@daiict.ac.in; arnab.kr311@gmail.com}
\affiliation{Dhirubhai Ambani Institute of Information and 
Communication Technology, Gandhinagar 382007, Gujarat, India}

\date{\today}

\begin{abstract}
We demonstrate the effectiveness of the logistic function to 
model the evolution of two economic systems. The first is the 
GDP and trade growth of the USA, and the second is the 
revenue and human resource growth of IBM. Our modelling 
is based on the World Bank data in the case of the 
USA, and on the company data in the case of IBM. 
The coupled dynamics of the two relevant variables in 
both systems --- GDP and trade for the  USA, and 
revenue and human resource for IBM --- 
follows a power-law behaviour. 
\end{abstract}



\maketitle

\section{The logistic function}
\label{sec1} 
The logistic equation is a standard example of a first-order
autonomous nonlinear dynamical system~\citep{stro}. 
Introduced originally to study population dynamics~\citep{stro,braun}, 
it was later applied to various problems of
socio-economic~\citep{braun,mon78,modis,akr10,kr22} and scientific
interest~\citep{stro}.
The growth of many natural systems 
is modelled accurately by the logistic equation,
the growth of species being a case in point~\citep{braun}. 
Modelling with the logistic equation is thus compatible with  
natural evolution itself.
This principle can be extended to the free evolution of
economic systems as well, a view that is supported 
by the successful logistic modelling of the GDP-trade dynamics 
of national 
economies~\citep{kr22} and industrial dynamics~\citep{akr10}.   

First-order autonomous dynamical systems have the general
form of $\dot{x} \equiv {\mathrm d}x/{\mathrm d}t = f(x)$ where
$x \equiv x(t)$, with $t$ being time~\citep{stro}.
Such a system may be linear or nonlinear, depending
on $f(x)$ being, respectively, a linear or a nonlinear
function of $x$~\citep{stro}. A basic model of a nonlinear
function is given by $f(x)=ax-bx^2$, with $a$ and $b$ being fixed
parameters. This leads to the well-known logistic equation,
\begin{equation} 
\label{logistic} 
\dot{x} \equiv \frac{{\mathrm d}x}{{\mathrm d}t} = f(x) = ax-bx^2. 
\end{equation}
Under the initial condition of $x(0)=x_0$, and with the definition
of $k=a/b$, the integral solution of Eq.~(\ref{logistic}) is
\begin{equation} 
\label{sologis} 
x(t) = \frac{kx_0 e^{at}}{k+x_0(e^{at}-1)},  
\end{equation}
which is the logistic function. 
From Eq.~(\ref{sologis}) we see that $x$ converges to the limiting
value of $k$ when $t \longrightarrow \infty$. This limit is known
as the carrying capacity in studies of population dynamics, and it
is also a fixed point of the dynamical system~\citep{stro}. This
becomes clear when we set the fixed point
condition $\dot{x}=f(x)=0$~\citep{stro}.
The two fixed points that result from Eq.~(\ref{logistic}) 
are $x=0$ and $x=k=a/b$.

On early time scales, when $t \ll a^{-1}$, the growth of $x$ can
be approximated to be exponential, i.e. $x \simeq x_0 \exp(at)$.
This gives $\ln x \sim at$, which is a linear relation on a
linear-log plot. Furthermore, we can interpret $a \simeq \dot{x}/x$
as the relative (or fractional) growth rate in the
early exponential regime.
However, this exponential growth is not indefinite,
and on times scales of $t \gg a^{-1}$ (or $t \longrightarrow \infty$)
there is a convergence to $x=k$. Clearly, the transition from the
exponential regime to the saturation
regime occurs when $t \sim a^{-1}$. This
time scale corresponds to the time
when the nonlinear term in Eq.~(\ref{logistic})
becomes significant compared to the linear term. The precise time for
the nonlinear effect to start asserting itself can be determined from
the condition $\ddot{x}= f^\prime (x) \dot{x} =0$ when $\dot{x} \neq 0$,
with the prime indicating a derivative with respect to $x$.
This requires solving $f^\prime (x) = a-2bx =0$ to get $x=a/2b =k/2$.
Using $x=k/2$ in Eq.~(\ref{sologis}) gives the nonlinear time scale as
\begin{equation} 
\label{nonlint}
t_{\mathrm{nl}} = \frac{1}{a} \ln \left(\frac{k}{x_0} -1\right), 
\end{equation}
which, we stress again, is the maximum duration over which a robust
exponential growth can be sustained.
Hereafter, we shall use Eqs.~(\ref{sologis})~and~(\ref{nonlint}) to
model two different economic systems. The first is the GDP-trade 
dynamics of the USA, whose national economy leads
the world. The second is the revenue and human 
resource growth of the company, IBM. 

\section{The coupled dynamics of GDP and trade}
\label{sec2} 
The GDP (Gross Domestic Product) of a country
is the market value of goods and services produced by the country
in a year~\citep{samnord,mas07,gmacl07}. GDP thus quantifies the
aggregate outcome of the economic activities of a country that are
performed all round the year. As such, the GDP of a national economy
is a dynamic quantity and its evolution (commonly implying growth)
can be followed through time. 

Contribution to the GDP of a country comes from another dynamic
quantity --- the annual trade in which the country engages
itself~\citep{gmacl07}.
The global trade network among countries exhibits some
typical properties of a complex network, namely, a scale-free
degree distribution and small-world clusters~\citep{sb03}.
If countries are to be treated as vertices in this network, then
global trade can be viewed as the exchange of wealth among the
vertices~\citep{gl04}.
The fitness of a vertex (a country) is measured by
its GDP, which also stands for the potential ability of a vertex
to grow trading relations with other vertices~\citep{gl04}.
Moreover, GDP itself follows its own power-law
distribution~\citep{gl04,gmacl07}, which in turn determines the topology
of the global trade network~\citep{gl04}. 
In qualitative terms,
these networks-based perspectives of the interrelation between GDP
and trade are in agreement with
the Gravity Model of trade, which mathematically formulates the trade
between two
countries to be proportional to the GDP of both~\citep{tin62}
(also see~\citep{jea10,dbt11} for subsequent reviews). 
Considering all of the foregoing facts together,
it is quite evident that GDP and trade
are intimately correlated. Both form a coupled system, in which the
dynamics of the one reinforces the dynamics of the other.

We look at the coupled dynamics of GDP and trade
within the mathematical framework of the logistic
equation~\citep{stro,braun}. This is in line with a 
study carried out on countries that are ranked high globally in terms
of their national GDPs~\citep{kr22}.
The temporal evolution of the total GDP
of the world economy (measured in US dollars) from 1870 to 2000 
does indicate a logistic trend~\citep{mas07}.
Empirical evidence also exists for a power-law feature in the
interdependent growth of GDP and trade~\cite{bmskm08}. 
We unify these observations in a theoretical model based on 
World Bank data that specifically pertain to the annual GDP 
and trade growth of the USA~\citep{usgd,ustd}.

\begin{figure}[]
\begin{center}
\includegraphics[scale=0.65, angle=0]{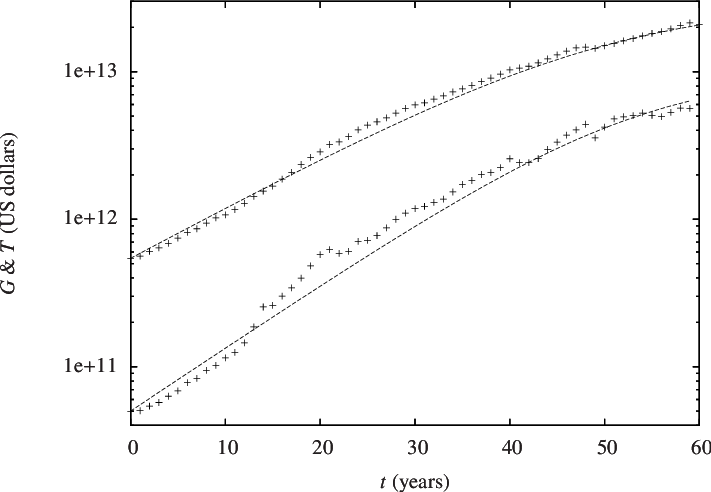}
\caption{\label{f1}\small{
Modelling the dynamics of GDP (the upper plot)
and trade (the lower plot) using World Bank data for 
the USA~\citep{usgd,ustd}. The zero year of both plots is 1960. 
The GDP plot ends in 2020, but the trade plot ends in 2019.
The two smooth dotted curves follow the logistic function, as 
given by Eq.~(\ref{sologis}). 
The parameter 
values to fit the GDP growth are $\gamma_1 = 0.080\,{\mathrm{year}^{-1}}$
and $k_G = \$\,30$ trillion (the predicted maximum value of the 
GDP). With 
respect to the logistic function, the yearly relative variation of the 
GDP data has a mean $\mu_G = 0.0492$ and a standard deviation  
$\sigma_G = 0.0873$. 
The parameter
values to fit the trade growth are $\tau_1 = 0.099\,{\mathrm{year}^{-1}}$
and $k_T = \$\,10$ trillion (the predicated maximum value of trade). With 
respect to the logistic function, the yearly relative variation of the
trade data has a mean $\mu_T = 0.1160$ and a standard deviation 
$\sigma_T = 0.2040$.
}}
\end{center}
\end{figure}

We quantify GDP by the variable $G \equiv G(t)$, with $G$
measured in US dollars and $t$ in years. To model
the annual growth of $G(t)$ with the logistic equation, as in
Eq.~(\ref{logistic}), we write
\begin{equation}
\label{gdplog} 
\dot{G} \equiv \frac{{\mathrm d}G}{{\mathrm d}t} = {\mathcal G}(G)
= \gamma_1 G - \gamma_2 G^2. 
\end{equation}
Noting that $x$, $a$ and $b$ in Eq.~(\ref{logistic}) translate,
respectively, to $G$, $\gamma_1$ and $\gamma_2$ in Eq.~(\ref{gdplog}),
we can write the integral solution of $G(t)$
in the same form as Eq.~(\ref{sologis}). It then follows that
when $t \longrightarrow \infty$, $G(t)$ converges to a limiting
value, i.e. $G \longrightarrow k_G = \gamma_1/\gamma_2$.
The early exponential growth of the GDP of the USA and its 
later convergence to a finite limit are modelled in the
upper linear-log plot in Fig.~\ref{f1}.
The smooth dotted curve tracks the GDP data~\citep{usgd} 
with the integral solution of Eq.~(\ref{gdplog}),
which will be in the mathematical form of Eq.~(\ref{sologis}).

As with GDP, the growth of trade can also be modelled 
with the logistic equation~\citep{kr22}. 
The annual trade of a country accounts for the total
import and export of goods and services.
The World Bank data on the annual trade of the USA 
are given as a percentage of the annual GDP~\citep{usgd,ustd}. 
Knowing the annual GDP, the trade percentage can be
expressed explicitly in terms of
US dollars, which we denote by the variable $T \equiv T(t)$,
with $t$ continuing to be measured in years.
We model the dynamics of $T(t)$ with the logistic equation,
as done in Eq.~(\ref{gdplog}), and write
\begin{equation}
\label{trdlog} 
\dot{T} \equiv \frac{{\mathrm d}T}{{\mathrm d}t} = {\mathcal T}(T)
= \tau_1 T - \tau_2 T^2. 
\end{equation}
Comparing Eq.~(\ref{trdlog}) with Eq.~(\ref{logistic}), we note
that $x$, $a$ and $b$ translate, respectively, to $T$,
$\tau_1$ and $\tau_2$. Hence, from the integral solution of $T(t)$,
which will be in the same form as Eq.~(\ref{sologis}), we will get
a convergence of $T \longrightarrow k_T = \tau_1/\tau_2$,
when $t \longrightarrow \infty$.
The integral solution of Eq.~(\ref{trdlog}) fits the 
trade data of the USA~\citep{ustd}
in the lower linear-log plot in Fig.~\ref{f1}.  

We note further in Fig.~\ref{f1} that the logistic modelling of
trade growth closely resembles the
logistic modelling of the GDP growth.
The similarity between the two plots is captured by 
a correlation coefficient of $0.992$ between the GDP and the 
trade of the USA~\citep{kr22}.
This high correlation is expected, because GDP and trade are dynamically 
connected to each other~\citep{gl04,gmacl07,mas07}.
As such, the coupled dynamics of GDP and trade must be governed by an
autonomous system of the second order, given as
${\dot T}={\mathcal T}(T,G)$ and
${\dot G} = {\mathcal G}(T,G)$. The
$T$-$G$ phase solutions are determined by integrating
\begin{equation} 
\label{phase} 
\frac{{\mathrm d}G}{{\mathrm d}T} = \frac{\dot{G}}{\dot{T}} = 
\frac{{\mathcal G}(T,G)}{{\mathcal T}(T,G)}
\end{equation}
for various initial values of the $(T,G)$ coordinates~\citep{stro}.
Since the 
functions ${\mathcal G}(T,G)$ and ${\mathcal T}(T,G)$ are not known
a priori, we apply a linear ansatz of
${\mathcal G} \simeq \gamma_1 G$ in Eq.~(\ref{gdplog}) and
${\mathcal T} \simeq \tau_1 T$ in Eq.~(\ref{trdlog}).\footnote{
For the coupled growth of $G$ and $T$,
a second-order dynamical system like ${\dot G} \sim T$ and
${\dot T} \sim G$ may appear apt. This, however, gives phase solutions
like $G^2 \sim T^2$, which is not borne out by a study of
GDP and trade growth~\citep{bmskm08}.
}
This approach agrees with the multiplicative character of
GDP and trade, whereby the revenue generated in one year is reinvested
in the economic cycle of the next year~\citep{gmacl07}.
The linearization gives a scaling formula 
(with $\alpha = \gamma_1/\tau_1$)
\begin{equation} 
\label{scale} 
G(T) \sim T^{\alpha}.
\end{equation}
Empirical evidence to support the power law implied by 
Eq.~(\ref{scale}) was found from 1948 to 2000, in a survey
of nearly two dozen countries of varying economic strength (high,
middle and low-income economies)~\citep{bmskm08}.

\begin{figure}[]
\begin{center}
\includegraphics[scale=0.65, angle=0]{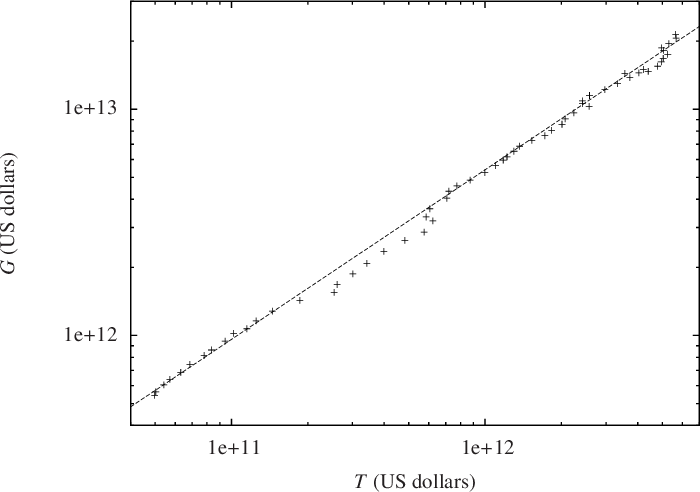}
\caption{\label{f2}\small{
Plotting GDP against trade using World Bank data for
the USA~\citep{usgd,ustd}. The plot begins in 1960 and
ends in 2019.
The straight dotted line follows the power-law function in
Eq.~(\ref{scale}) with
$\alpha = 0.75$, a value that is close to
$\gamma_1/\tau_1 \simeq 0.81$ (as given in Fig.~\ref{f1}).
With respect to the logarithm of the power-law function, the
yearly variation of the logarithm of the GDP data has a mean
$\mu_\alpha = -0.0012$ and a standard deviation 
$\sigma_\alpha = 0.0024$.
}}
\end{center}
\end{figure}

The power-law function in Eq.~(\ref{scale}) becomes linear in a 
log-log plot. This is indeed
what we see in Fig.~\ref{f2} which models the
coupled growth of the GDP and trade of the USA. 
The power-law exponent $\alpha$ is given by the slope of the linear 
fit, within the range of $0 < \alpha < 1$~\citep{kr22}. 
Keeping only the linear
terms in Eqs.~(\ref{gdplog})~and~(\ref{trdlog}), which lead to the
phase solutions in Eq.~(\ref{scale}),
we find that $\alpha = \gamma_1/\tau_1$. The values of
$\gamma_1$, $\tau_1$ and $\alpha$, required for plotting 
Figs.~\ref{f1}~and~\ref{f2}, do show that
$\alpha$ is quite close to $\gamma_1/\tau_1$. This
independently validates our modelling of GDP and trade growth with
the logistic equation.

Looking at Fig.~\ref{f2}, we realize that the power-law
scaling of $G$ with respect to $T$ holds true over
nearly three orders of magnitude.
For high values of $T$ and $G$, 
deviation from this scaling behaviour is possible 
due to the nonlinear effects in the
real data~\citep{kr22}. However, we have not considered nonlinearity 
in the coupled autonomous
functions ${\mathcal G}(T,G)$ and ${\mathcal T}(T,G)$
to derive Eq.~(\ref{scale})~\citep{kr22}. We also note that
${\mathrm d}^2 G/{\mathrm d}T^2 < 0$ for $\alpha < 1$, i.e. $G$
increases with $T$ at a decreasing rate as time progresses.
This explains the steady reduction of
the gap between the GDP and the trade plots in Fig.~\ref{f1}
on long time scales.

\section{The logistic dynamics of a company}
\label{sec3}
The growth (and the health) of a company is gauged by the annual 
revenue that it generates and the human resource that it employs.
Regular monitoring of these two variables is necessary
for a precise understanding of the patterns of industrial growth. 
Even when a company shows noticeable growth in the early 
stages, a saturation in its growth occurs on later 
time scales~\citep{aghow}. 
Clearly, as the size of an organization increases, its growth 
rate becomes progressively inhibited. 
Therefore, to explain saturation in industrial growth,  
an effective mathematical model has to study 
the growth of a company that operates on the largest possible 
scale, which can only be the global scale. 
What is more, when a company operates on the global scale, its
overall growth pattern becomes free of local inhomogeneities.
This itself affords an advantage for the mathematical modelling. 
In view of this, we analyze the growth of the annual revenue
and the human resource strength of the multi-national
company, IBM. 
Data about its annual revenue generation, the net annual earnings
and the cumulative human resource strength,
dating from 1914 to 2006, are available on the company
website~\citep{ibmhis}. 
\begin{figure}[]
\begin{center}
\includegraphics[scale=0.65, angle=0]{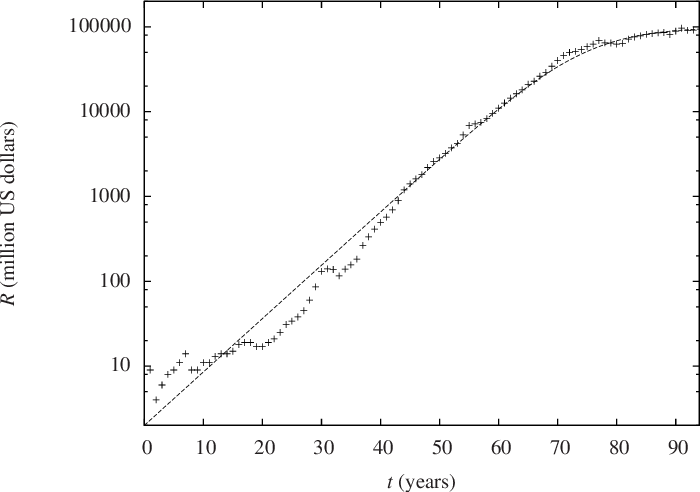}
\caption{\label{f3}\small{
Modelling the annual revenue growth of IBM, using 
the company data from 1914 to 2006~\citep{ibmhis}. 
The smooth dotted curve is the logistic function, 
as given by Eq.~(\ref{sologis}). The parameter values to fit 
the revenue growth are $\rho_1 = 0.145\,{\mathrm{year}^{-1}}$
and $k_R = \$\,100$ billion (the predicted maximum revenue that
IBM can earn). With
respect to the logistic function, the yearly relative variation 
of the revenue has a mean $\mu_R = 0.025$ and a standard 
deviation $\sigma_R = 0.4897$. Saturation of the revenue growth 
starts on the time scale of $75$-$80$ years.
}}
\end{center}
\end{figure}

As we have done for the GDP and trade dynamics of the USA 
in Sec.~\ref{sec2}, 
we posit the logistic equation to model 
the annual revenue and human resource growth of IBM. 
If the revenue is $R \equiv R(t)$, with $R$
measured in US dollars and $t$ in years, then the logistic model 
for the revenue growth is  
\begin{equation}
\label{revlog} 
\dot{R} \equiv \frac{{\mathrm d}R}{{\mathrm d}t} 
= {\mathcal R}(R) = \rho_1 R - \rho_2 R^2.   
\end{equation}
Since $x$, $a$ and $b$ in Eq.~(\ref{logistic}) 
translate, respectively, 
to $R$, $\rho_1$ and $\rho_2$ in Eq.~(\ref{revlog}),
the integral solution for $R(t)$ will be 
in the same form as Eq.~(\ref{sologis}). And so 
when $t \longrightarrow \infty$, $R(t)$ will converge to a limiting
value of $R \longrightarrow k_R = \rho_1/\rho_2$.
The early exponential growth of the revenue of IBM and its
later saturation to a finite limit are modelled in the
linear-log plot in Fig.~\ref{f3}.
The smooth dotted curve fits the revenue data~\citep{ibmhis}
according to the logistic function, in the 
form of Eq.~(\ref{sologis}). The most noteworthy feature in Fig.~\ref{f3} 
is the saturation in the revenue growth of IBM around $75$-$80$ years. 
This time scale can be also obtained from the formula of the nonlinear 
time scale in Eq.~(\ref{nonlint}), by equating the parameter values 
as $a = \rho_1$ and $k=k_R$. From this it is clear that the revenue
growth of IBM entered the nonlinear regime around the time of 
$75$-$80$ years (the initial years of the 1990s). 

\begin{figure}[]
\begin{center}
\includegraphics[scale=0.65, angle=0]{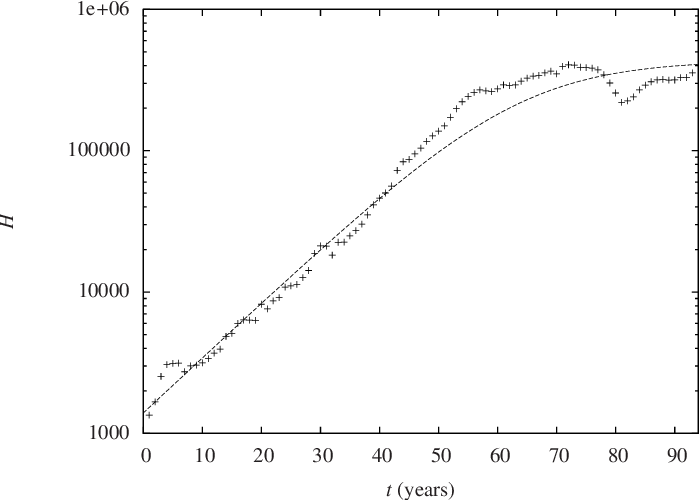}
\caption{\label{f4}\small{
Modelling the human resource growth of IBM, using
the company data from 1914 to 2006~\citep{ibmhis}.
The smooth dotted curve is the logistic function,
as given by Eq.~(\ref{sologis}). The parameter values to fit
the human resource growth are $\eta_1 = 0.09\,{\mathrm{year}^{-1}}$
and $k_H = 500000$ (the predicted maximum employees of IBM). 
With respect to the logistic function, the yearly relative variation
of the human resource has a mean $\mu_H = 0.106$ and a standard
deviation $\sigma_H = 0.2999$. The human 
resource graph declines around $75$-$80$ years.  
}}
\end{center}
\end{figure}

Now we write the human resource of IBM as 
$H \equiv H(t)$, with $t$ in years as usual.  
Then translating $x$, $a$ and $b$ in Eq.~(\ref{logistic}),
respectively, to $H$, $\eta_1$ and $\eta_2$,  
the logistic equation for the human resource growth becomes 
\begin{equation}
\label{hrlog} 
\dot{H} \equiv \frac{{\mathrm d}H}{{\mathrm d}t} 
= {\mathcal H}(H) = \eta_1 H - \eta_2 H^2, 
\end{equation}
whose integral solution will have the 
form of Eq.~(\ref{sologis}). 
With $t \longrightarrow \infty$, a convergence to a limiting
value occurs for $H(t)$, which goes 
as $H \longrightarrow k_H = \eta_1/\eta_2$.
The early exponential growth of the human resource of IBM  
and its later convergence to a finite limit are modelled 
in the linear-log plot in Fig.~\ref{f4}.
The smooth dotted curve is the model logistic function and it 
follows the human resource growth of IBM~\citep{usgd}. 

\begin{figure}[]
\begin{center}
\includegraphics[scale=0.65, angle=0]{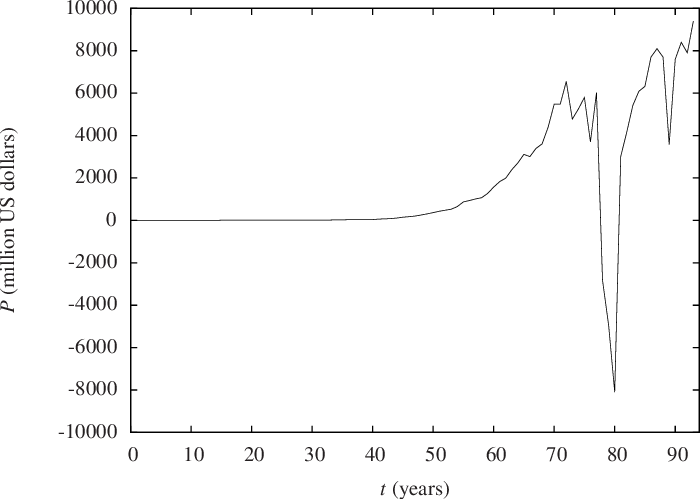}
\caption{\label{f5}\small{
The net annual earnings (the profit $P$) of IBM grow steadily  
till about
$75$-$80$ years (the early years of the 1990s). Around this time 
IBM suffered major losses in its net earnings 
($\$\,8$ billion in 1993), and this time scale 
corresponds closely to the time scale for the onset of nonlinear 
saturation in revenue growth, which is also $75$-$80$ years.
}}
\end{center}
\end{figure}

A point to note in Fig.~\ref{f4}
is the depletion of the human resource of IBM on the time scale of
$75$-$80$ years. This is evidently correlated with the saturation
of the revenue growth of IBM on the same time scale, as
Fig.~\ref{f3} shows. That the revenue and the human resource of a
company are correlated is entirely to be expected.
If a company generates enough revenue, it becomes financially
viable for it to maintain a functioning human resource pool,
which in turn generates further revenue. In
this manner both the revenue and the human resource of a company
sustain the growth of each other. Whenever one of the variables
is affected adversely, there is an equally adverse impact on the 
other variable.
In the case of IBM, the saturation of its revenue growth around
$75$-$80$ years resulted in a loss of human resource on the same
time scale. An additional confirmation of this argument comes from 
Fig.~\ref{f5}, 
which plots the net annual earnings (the profit $P$) of IBM 
against time.
The company suffered major financial losses
in the early 1990s (upto $\$\,8$ billion in 1993), which 
matches our estimate of the nonlinear time scale of $75$-$80$ years.  
\begin{figure}[]
\begin{center}
\includegraphics[scale=0.65, angle=0]{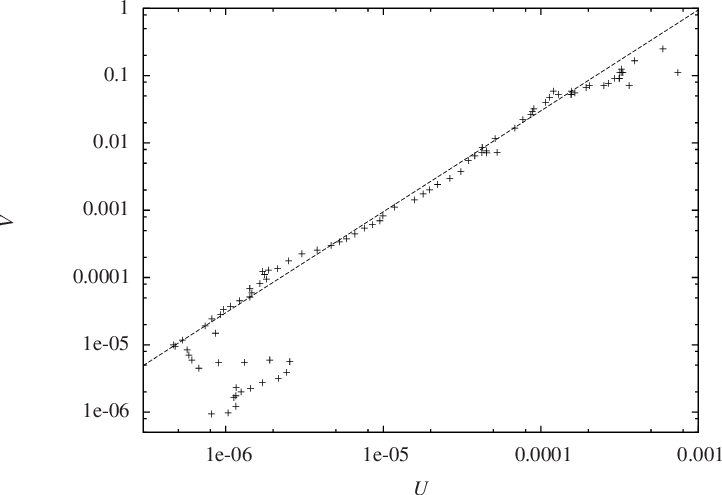}
\caption{\label{f6}\small{
Fitting Eq.~(\ref{uvpow}) to the correlated growth of $H$ and $R$, 
with $\beta = 1.5$ (close to  
$\beta = \rho_1/\eta_1 \simeq 1.6$). 
The cusp in the data points at the bottom left is due to  
human resource loss around $75$-$80$ years.
}}
\end{center}
\end{figure}

To analyze the correlated growth of $R$ and $H$, we set down a 
coupled autonomous dynamical system as 
$\dot{R} = {\mathcal R}(H,R)$ and $\dot{H} = {\mathcal H}(H,R)$.
As in the case of the derivation of Eq.~(\ref{scale}), the 
coupled autonomous functions ${\mathcal R}(H,R)$ and 
${\mathcal H}(H,R)$ are not known a priori. 
Therefore, in a basic approach, we assign to these functions the 
uncoupled logistic forms in Eqs.~(\ref{revlog})~and~(\ref{hrlog}), 
respectively. This simplifies to 
$\dot{R} = {\mathcal R}(R)$ and $\dot{H} = {\mathcal H}(H)$. The 
variable $t$, which is implicit in this set of equations, can be 
eliminated to obtain the $H$-$R$ phase solutions for initial 
values of the $(H,R)$ coordinates~\citep{stro}. 
Defining $V = R^{-1}- k_R^{-1}$, $U = H^{-1}- k_H^{-1}$ and
$\beta = \rho_1/\eta_1$,
the $H$-$R$ phase solutions are transformed to a compact power-law 
form as 
\begin{equation} 
\label{uvpow} 
V \sim U^\beta. 
\end{equation} 
The power-law in Eq.~(\ref{uvpow}) fits the data well in the 
log-log plot in Fig.~\ref{f6}, except for the cusp at 
the bottom left. 
However, the lower arm of the cusp has nearly the same positive 
slope as the extended straight-line fit in Fig.~\ref{f6}, which shows 
that intermittent deviations do not affect the overall 
growth too much~\cite{mon78}.

\section{Concluding remarks}
\label{sec4} 
The fact that the logistic equation shows a saturation in growth 
on long time scales implies 
long-term economic stagnation.
Reasons for this are dwindling natural resources, natural calamities,
pandemics, obsolescence of technology, military conflicts, etc. The
decisive reasons are often unforeseen. Nevertheless, the logistic
equation continues to be a favoured mathematical tool for modelling
the evolution of socio-economic systems~\citep{braun,mon78}. For
example, our use of the logistic equation and the
power-law correlation function in the phase plot was equally
effective in modelling the GDP-trade dynamics of six top 
national economies at present~\citep{kr22} and 
industrial growth~\citep{akr10}.
This analogy between national economies and companies is of
interest because studies point to universal mechanisms that
underlie the economic dynamics of countries and
companies~\citep{sabhlmss96,lacms98}.
This commonality can help in understanding the dynamics of
large companies, whose stock values grow
to the scale of national economies. 

The economy of the USA is suited well for our logistic modelling 
because of its balanced GDP growth,  
as we can see from the
closeness between the theoretical logistic function and the actual
GDP data in Figs.~\ref{f1}~and~\ref{f2}.
Moreover, from a macroeconomic perspective, GDP
is a reliable yardstick of the state of a national economy,
and in a global comparison of national economies,
the USA has the highest GDP in the world at present. The  
balanced and robust growth of the US economy has been possible 
because of democratic values
in internal politics, the absence of military conflicts on
the national borders, and the promotion of free economic 
growth~\citep{kr22}. 

Global economic recessions from time to time make it imperative 
to devise accurate mathematical models for understanding
economic stagnation and for predicting correct outcomes. 
The logistic equation proves effective in both respects, 
as has been demonstrated by a recent study on the GDP competitiveness 
among some leading national economies~\citep{akr23}. 
That said, we also have to remember that 
studies of this type come
under the general category of social systems, 
and consequently their predictive power
depends on socio-economic factors.  
Unforeseen natural, social and political events can compromise 
the forecasts made in these studies, and force mathematical 
models to be recalibrated.  


\bibliography{lm}
\end{document}